\documentclass[12pt]{article}
\usepackage{amssymb,graphicx}
\usepackage{amsmath}
\usepackage{epsfig}

\def\e{{\rm e}}

\def\l{\left(}
\def\r{\right)}

\newcommand{\be}{\begin{equation}}
\newcommand{\ee}{\end{equation}}
\newcommand{\bea}{\begin{eqnarray}}
\newcommand{\eea}{\end{eqnarray}}
\newcommand{\bg}{\begin{gather}}
\newcommand{\eg}{\end{gather}}
\newcommand{\bseq}{\begin{subequations}}
\newcommand{\eseq}{\end{subequations}}

\renewcommand{\tanh}{\mathop{\rm th}\nolimits}
\newcommand{\ch}{\mathop{\rm ch}\nolimits}
\renewcommand{\ln}{\mathop{\rm ln}\nolimits}

\newcommand{\Tr}{{\rm Tr}}

\begin{document}
\begin{center}
  {\Large\bf Domain walls in noncommutative gauge theories, folded
 D-branes, and communication with mirror world} \\
\medskip
S.L.~Dubovsky$^{a,b}$, S.M.~Sibiryakov$^b$\\
\medskip
  $^a${\small
Department of Physics, CERN Theory Division CH-1211 Geneva 23, Switzerland
  }\\
\medskip
$^b${\small
Institute for Nuclear Research of
         the Russian Academy of Sciences,\\  60th October Anniversary
  Prospect, 7a, 117312 Moscow, Russia
  }

\end{center}

\begin{abstract}
Noncommutative $U({\cal N})$ gauge theories at different ${\cal N}$
may be often thought
of as different sectors of a single theory. For instance,
$U(1)$ theory possesses
a sequence of vacua labeled by an integer parameter ${\cal N}$, and the theory
in the vicinity of the ${\cal N}$-th vacuum coincides with the $U({\cal N})$
noncommutative gauge theory. We construct domain walls
on noncommutative plane, which separate
vacua with different gauge groups in gauge
theory with adjoint scalar field.
The scalar field has 
nonminimal coupling to the gauge field, such that the scale of
noncommutativity is determined by the vacuum value of the scalar field.
The domain walls are
solutions of the BPS equations in the theory.
It is natural to interprete the domain wall as a stack of
D-branes plus a stack of folded D-branes. We support this
interpretation by the analysis of small fluctuations around domain
walls, and suggest that such configurations of branes emerge as
solutions of the Matrix model in large class of pp-wave backgrounds with
inhomogeneous field strength.
We point out that the folded D-brane per se provides an explicit realization of the
``mirror world'' idea, and speculate on some phenomenological
consequences of this scenario.
\end{abstract}
\section{Introduction}
\label{intro}

Recent interest in noncommutative (NC) gauge theories is to large
extent motivated by rich pattern of stringy phenomena exhibited
in these theories.
One of the spectacular manifestations of the stringy
origin of NC gauge theories is that in many
cases~\cite{Gross:2000ss,Bak:2000zm,Bak:2001kq,Demidov:2003xq}
the rank ${\cal N}$ of gauge group is a background dependent parameter.
For
instance, it was pointed out in Ref.~\cite{Gross:2000ss} that $U(1)$
gauge theory on NC plane possesses a sequence of vacua labeled by a
natural number ${\cal N}$ with the following peculiar properties:
\newline
i) Every vacuum with ${\cal N}>1$ is a highly
non-local field configuration from the point of view of the trivial
(${\cal N}=1$) vacuum;
\newline
ii) Perturbation theory in the
vicinity of the ${\cal N}$-th vacuum is equivalent to perturbation
theory of the $U({\cal N})$ NC gauge theory above its
trivial vacuum;
\newline
iii) The fact that there are different gauge
theories in different vacua cannot be understood as Higgs
mechanism. Namely, the action in
the vicinity of the ${\cal N}$-th vacuum contains $U({\cal N})$ gauge
fields only, with no extra massive vector bosons or Higgs fields.

A possible interpretation of this phenomenon is that $U({\cal N})$
NC gauge theory
``remembers'' its origin in string theory, where it emerges as
the effective field theory describing a stack of ${\cal N}$ D-branes
with constant $B$-field in the zero slope limit~\cite{Seiberg:1999vs}
and, thus, the rank of gauge group ${\cal N}$
is a parameter of the background.
However, in the case of pure gauge theory on the NC plane it was
argued~\cite{Gross:2000ss} that ${\cal N}$ is a superselection
parameter, implying that different sectors are completely disconnected
from each other. Were this the whole story, one could object that in
this respect NC theories are not more ``stringy'' than
ordinary Yang--Mills theories. Indeed, one may think that the very
existence
of ordinary $U({\cal N})$ theories with different ${\cal N}$ is
also a reflection of
the fact that ordinary gauge theory appears as the low energy
description of a stack of ${\cal N}$ D-branes in string theory.

However, recently it was demonstrated that in some NC theories,
it is possible to construct domain walls
\cite{Dubovsky:2003ga} (see also Ref.~\cite{Bak:2001gm} for earlier
related work) interpolating between vacua with different gauge
groups. Consequently, the rank of the gauge group can be a
nontrivial {\it dynamical} parameter in NC gauge theories.
This makes the above phenomenon significantly more interesting from
the viewpoint of both field and string theory.

The NC manifold considered in Ref.~\cite{Dubovsky:2003ga} is fuzzy
culinder. Although the fuzzy cylinder setup enables one to construct domain
walls interpolating between vacua with
$U({\cal N})$ and $U({\cal M})$ gauge groups for arbitrary ${\cal N}$
and ${\cal M}$, it has a number of drawbacks. First, the algebra of
functions on fuzzy cylinder is different from that on the NC plane.
As a result, some formulae become more complicated and obscure.
Another, less technical problem is that
fuzzy cylinder can be interpreted as a
cylindrical semi-lattice. This may give rise to a suspicion, that it is the
lattice structure which is crucial for the existence of the domain
walls, and that they are absent in a continuous setting. This
suspicion is further corroborated by the fact that the tension of
walls constructed in Ref.~\cite{Dubovsky:2003ga} diverges in the limit
when the fuzzy cylinder approaches NC plane.

\begin{figure}[tb]
\begin{center}
\begin{picture}(500,100)(0,0)
\put(0,20){\epsfig{file=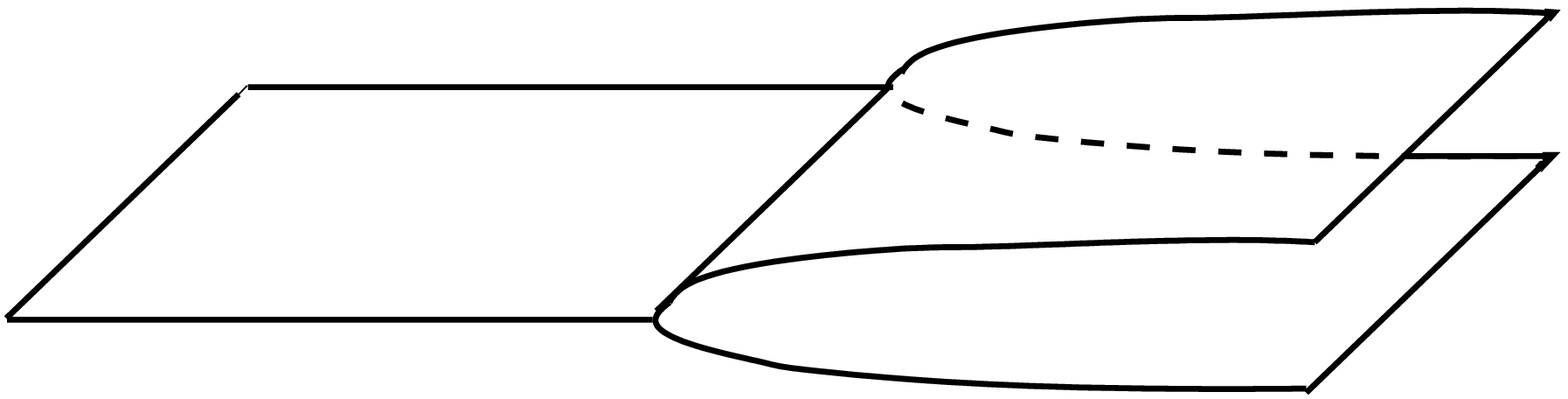,height=1.5cm,width=5.cm} }
\put(160,20){\epsfig{file=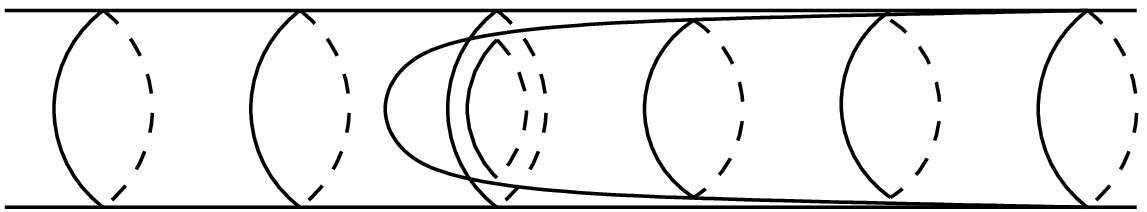,height=1.5cm,width=5.cm}}
\put(325,10){\epsfig{file=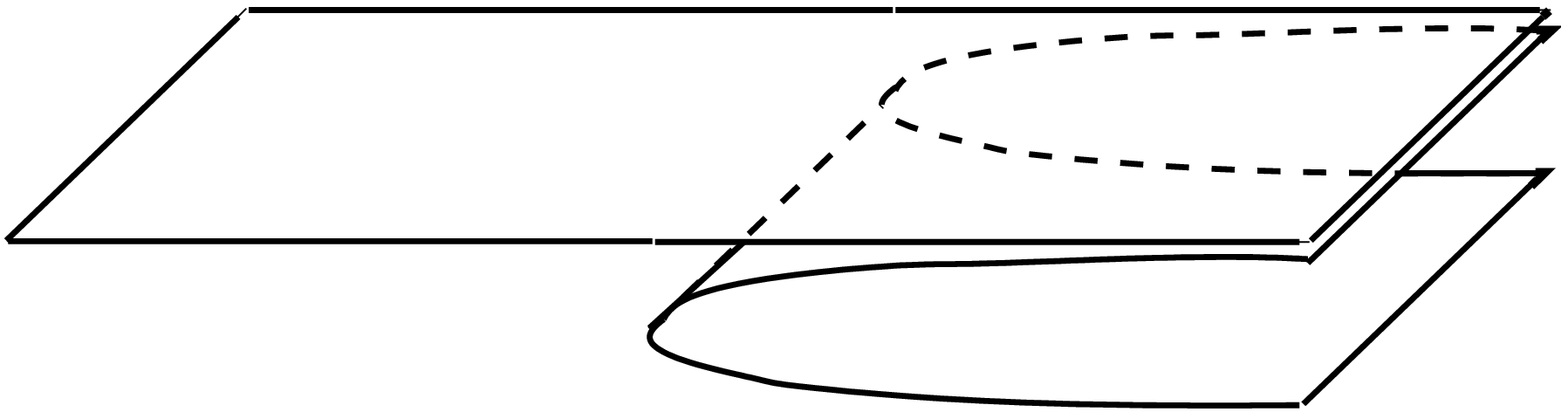,height=1.5cm,width=5.cm}}
\put(55,0){a}
\put(225,0){b}
\put(394,0){c}
\end{picture}
\end{center}
\caption{
a) hypothetical junction of D-branes, b) geometrical interpretation
of a domain wall on the fuzzy cylinder, c) geometrical
interpretation of a domain wall on NC plane.
}\label{walls}
\end{figure}

The main purpose of this paper is to demonstrate that
domain walls between vacua with different gauge groups exist also
in theories on the NC plane without any lattice structure.
According to Ref.~\cite{Gross:2000ss}, these theories cannot be pure
Yang-Mills. 
To understand the ingredients which should be added,
it is instructive to think about possible string
theoretical interpretation of the domain walls. The first idea
may be that the
domain wall is a kind of a junction of D-branes. For instance, in
Fig.~\ref{walls}a, we show the hypothetical
junction which would correspond
to a domain wall between $U(1)$ and $U(2)$
vacua. However, junctions of this type do not exist in string
theory. The geometrical interpretation of the domain walls
on the fuzzy cylinder suggests a different picture. Namely,
these domain walls can be understood as a set of
infinite cylindrical branes and half-infinite branes of fingerstall
shape \cite{Dubovsky:2003ga}. In Fig.~\ref{walls}b
the brane configuration
corresponding to $U(1)-U(2)$ domain wall on the fuzzy cylinder is
shown. A natural analogue of this configuration in the planar
case is shown in Fig.~\ref{walls}c. In the rest of the paper we
describe domain walls on NC plane and demonstrate
that their properties match the ones expected for the D-brane
configuration of Fig.~\ref{walls}c.

The first ingredient needed to construct a realization of the brane
configuration  of Fig.~\ref{walls}c in NC gauge theory
is extra scalar
field $\phi$ in the adjoint representation of the gauge group. This field
corresponds to the transverse coordinate in Fig.~\ref{walls}c and
parametrizes the relative positions of D-branes in
the transverse direction. Scalar fields of this type are generic in
gauge theories residing on stacks of D-branes both in commutative
and noncommutative cases.

Another modification of pure NC Yang--Mills theory needed for
incorporating  
domain walls is less trivial. It stems from the fact that the two sheets of
the folded brane in Fig.~\ref{walls}c appear as a brane-antibrane
pair far from the tip of the fold (and not as a pair of two D-branes).
A simple way to understand this is to consider the folded brane as
the limiting configuration obtained from the cylindrical brane by
elliptic deformation (c.f. Refs.~\cite{Bak:2001xx,Bak:2002wy}).
As the cylindrical brane carries no net
RR-charge, two sheets of the fold should carry opposite charges.
To figure out what is needed to satisfy this requirement in the
NC theory, it is helpful to use the language of the
BFSS Matrix model~\cite{BFSS}. There, the RR-charge density in the
transverse plane $(X,Y)$ is proportional to the commutator of the
corresponding matrix coordinates,
\[
\rho_{XY}\propto i[X,Y]\;,
\]
and this charge density should take opposite values on the two sheets
of the folded brane.
On the other hand, in the NC theory the commutator of two coordinates
is the parameter of the noncommutativity,
\[
[X,Y]=i\theta\;.
\]
Consequently, one should allow the parameter of noncommutativity
to vary in space, so that it takes opposite values at the two sheets
of the brane. We suggest that this variation comes from
nonminimal (gauge invariant) interaction of the adjoint scalar
field with gauge fields. As a result of this interaction, the
parameter of noncommutativity is actually determined by the VEV of the
scalar field (for a similar construction of supersymmetric brane-antibrane
configurations see Refs.~\cite{Bak:2001xx,Bak:2001tt,Bak:2002wy}). 
Then, after the scalar
potential of the double well shape is added, the resulting theory
has a set of vacua with $U({\cal N})\times U({\cal M})$ gauge
groups. These vacua
correspond to stacks of ${\cal N}$ branes
and ${\cal M}$ antibanes. Also, there are BPS domain walls interpolating
between $U({\cal N})\times U({\cal M})$ and
$U({\cal N}+{\cal K})\times U({\cal M}+{\cal K})$
vacua for any ${\cal N}$, ${\cal M}$ and ${\cal K}$.

The rest of the paper is organized as follows. In section \ref{model}
we describe our setup and study the basic properties of the $U({\cal
N})\times U({\cal M})$ vacua. In section \ref{WALL} we derive the BPS
bound for the energy of the domain wall and solve the corresponding
BPS equations. In section \ref{Adjoint} we introduce the probe adjoint
scalar field and study its spectrum in the domain wall
background. This analysis supports our interperetation of the domain
wall as a brane configuration shown in Fig.~\ref{walls}c. In
the concluding section \ref{final} we summarize our results,
discuss the relation between our  model
and Matrix theory in curved background
and speculate on some implications of our results for the brane
world/mirror world scenario.

\section{The model}
\label{model}
Let us follow the steps oulined in the Introduction and
construct the action of gauge theory on NC plane admitting
domain walls between vacua with different gauge
groups.
To start with, let us consider the action of the pure NC $U(1)$
theory,
\be
\label{YM}
S_{YM}=-\frac{1}{4g^2}\int d^3x\;F_{\lambda\nu}^2\;,
\ee
where $\lambda,\nu=t,x,y$.
This action has the same form as that of the ordinary Yang--Mills
theory. The only difference is that the Moyal $*$-product
with the
space-space noncommutativity,
\[
[x,y]_*=i\theta
\]
replaces the ordinary product of fields everywhere.
Now, let us introduce a real scalar field $\phi$ in the adjoint
representation of the gauge group with the double well potential
\be
\label{scalar}
S_{\phi}={1\over g^2}\int d^3x\left(\frac{1}{2}(D_{\nu}\phi)^2-{\mu^2\over
2}\l\phi^2-v^2\r^2\right)\;,
\ee
where again the $*$-product is assumed everywhere.

The next step is to add extra terms which will ensure that the
noncommutativity parameter effectively changes its sign in the vacuum
with $\phi=-v$. To this end, let us perform the Weyl map and write the
gauge field action (\ref{YM}) in the operator language (see, e.g.
Refs.~\cite{harvey} for details),
\be
\label{YMop}
S_{YM}=\int dt\, \frac{2\pi\theta}{g^2}\, \Tr\l
\frac{1}{2\theta^2}(D_0 X)^2 +\frac{1}{2\theta^2}(D_0 Y)^2
-\frac{1}{2\theta^4}\left(i[X,Y]+\theta\right)^2\r\;,
\ee
where covariant coordinates $X, Y$ are related to the gauge
potentials as follows,
\begin{align}
X&=\hat{x}+\theta \hat{A}_y
\label{Ay}\\
Y&=\hat{y}-\theta \hat{A}_x
\label{Ax}\;.
\end{align}
Hats mean that we are working with operators in a Hilbert
space.
The $(x,y)$-component of the field strength tensor is equal to
\be
\label{Fxy}
F_{xy}=-{i\over\theta^2}[X,Y]-{1\over \theta}\;.
\ee
In the operator formulation the only place where the sign of $\theta$
comes in is the last term in Eq.~(\ref{YMop}). To change this sign in
the vacuum with $\phi=-v$ we introduce the extra $\phi$-dependence
into this term, by multiplying $\theta$ by a factor $\phi/v$. Then the
full action of the model takes the following form in the operator language
\begin{align}
\label{Act}
S=\int dt\, \frac{2\pi\theta}{g^2}\, \Tr\Bigl\{&
\frac{1}{2\theta^2}(D_0 X)^2 +\frac{1}{2\theta^2}(D_0 Y)^2
+\frac{1}{2}(D_0 \phi)^2\\
-&\frac{1}{2\theta^4}\left(i[X,Y]+\frac{\theta}{v}\phi\right)^2
-\frac{1}{2\theta^2}(i[X,\phi])^2 -\frac{1}{2\theta^2}(i[Y,\phi])^2
-\frac{\mu^2}{2}\left(\phi^2-v^2\right)^2\Bigr\}\nonumber
\end{align}
As we discuss in more detail in section \ref{final}, in the spirit of
Matrix model this action is naturally
interpreted as describing dynamics of
D0-branes in (curved) (3+1)-dimensional space-time. In this
interpretation, operators $X$, $Y$ and
\[
Z\equiv\theta\phi
\]
represent spatial coordinates of D0-branes.


In what follows we work in the gauge $A_0=0$.
Let us first classify time-independent vacuum solutions of the model.
It follows
immediately from Eq.~(\ref{Act})
that the energy is minimum (zero), provided that
\bseq
\label{vac}
\begin{gather}
\phi^2=v^2\\
[X,\phi]=[Y,\phi]=0\\
[X,Y]=i\frac{\theta}{v}\phi\;.
\end{gather}
\eseq
Solutions to Eqs. (\ref{vac}) are direct sums of any number
of two basic representations defined on the Hilbert space of
functions of one variable $q$,
\bseq
\label{vacrep}
\begin{gather}
\phi=v~,~~~X=\hat{q}~,~~~Y=-i\theta\frac{\partial}{\partial
  q}\label{Dbrane}\\
\phi=-v~,~~~X=-\hat{q}~,~~~Y=-i\theta\frac{\partial}{\partial
  q}\label{Dantibrane}\;.
\end{gather}
\eseq
From the Matrix model viewpoint,
a direct sum of ${\cal N}$ representations
of type (\ref{Dbrane}) and
${\cal M}$ representations of type (\ref{Dantibrane}) (${\cal N}-{\cal
M}$ vacuum)
corresponds to a
system of ${\cal N}$ D2-branes and ${\cal M}$ $\bar{\mbox{D}}2$-branes.
The separation between D-branes and $\bar{\mbox{D}}$-branes is given
by
\be
\label{a}
a=\theta v
\ee
The fact that under certain circumstances the brane-antibrane systems may
exhibit no instability was observed in a similar context in
Refs.~\cite{Bak:2001xx,Bak:2001tt,Bak:2002wy}.

It is worth noting that there exist various
space-time formulations of the action (\ref{Act}).
In particular, the original field-operator map
used to rewrite $U(1)$ gauge theory (\ref{YM}) in the operator
form (\ref{YMop}), translates the action (\ref{Act}) into the following field
theory action,
\be
\label{commut}
S=\int dtdxdy~\frac{1}{g^2}\Bigl(-\frac{1}{4}F_{\lambda\nu}^2
+\frac{1}{2}(D_{\nu}\phi)^2 +\frac{1}{a}F_{xy}(\phi-v)-{1\over
2a^2}\l\phi-v\r^2
-\frac{\mu^2}{2}\left(\phi^2-v^2\right)^2
\Bigr)\;,
\ee
where,
again, $*$-product is assumed everywhere. This space-time formulation
is well suited for the study of the single D-brane
vacuum
(\ref{Dbrane}), which corresponds to ${\cal N}=1$, ${\cal M}=0$.
Indeed, at the level of quadratic action the $*$-product is
equivalent to the ordinary product, and the action (\ref{commut}) describes
just a system of one vector and one scalar field with a certain
(Lorentz-violating)
mixing between them. Gauge field is zero in the vacuum (\ref{Dbrane})
and the VEV of the scalar field is equal to $v$.
There are two propagating degrees of freedom in this vacuum with the
following dispersion relations,
\be
\label{dispAphi}
\omega_{1,2}^2={k}^2+\frac{m_{\phi}^2}{2}\pm
\sqrt{\frac{m_{\phi}^4}{4}+ \frac{{ k}^2}{a^2}}
\ee
where $\omega$ is energy, ${k}$ is the absolute value of the momentum, and
\begin{gather}
\label{mphi}
m_{\phi}^2=\left(\frac{1}{\theta^2 v^2}+4\mu^2 v^2\right)\;.
\end{gather}
At low momenta, $k\ll a m_\phi^2$, Eq.~(\ref{dispAphi}) reduces to
\bseq
\label{dispAphi1}
\begin{align}
&\omega_1^2=m_{\phi}^2+{k}^2\left(1+\frac{1}{m_{\phi}^2
    a^2}\right)\\
&\omega_2^2={k}^2\left(1-\frac{1}{m_{\phi}^2 a^2}\right)\;.
\end{align}
\eseq
We see that Lorentz-violating effects are small at low energies
provided that
\be
\label{mutvcond}
\mu\theta v^2\gg 1
\ee
and the dispersion relations (\ref{dispAphi1}) describe one massless
particle
(photon) and one
massive particle (scalar boson) in this limit. Note that the
condition (\ref{mutvcond}) is compatible with weak coupling regime
in the naive
commutative limit $\theta\to 0$ at small enough gauge coupling
constant $g$ (we do not consider possible UV-IR mixing
effects here).
In what
follows it is the limit
\be
\label{commutlimit}
\theta\to 0\;,\;\; \mu\theta v^2\to\infty
\ee
which we refer to as the commutative limit.

From Eq.~(\ref{Fxy}) we see that
in the space-time formulation (\ref{commut}), $\bar{\mbox{D}}$-brane
vacuum (${\cal N}=0$, ${\cal M}=1$)
corresponds to the vacuum state with $\phi=-v$ and with
non-zero magnetic field $F_{xy}=-2/\theta$. On the other hand, in the
operator language it is clear that physics in this vacuum is identical
to physics in the ${\mbox{D}}$-brane vacuum. To see this
explicitly one may replace the ordinary Weyl ordering by a different
field-operator correspondence, defined as
\[
\hat{f}\leftrightarrow f(x,y)\equiv f_W(-x,y)\;.
\]
where $f_W(x,y)$ is the Weyl symbol of operator
$\hat f$. Also one changes relations (\ref{Ay}), (\ref{Ax})
between operators $X$, $Y$ and gauge fields, so that
\begin{align}
X&=-\hat{x}-\theta \hat{A}_y
\nonumber\\
Y&=\hat{y}+\theta \hat{A}_x
\nonumber
\;.
\end{align}
Then it is straightforward to check, that in the space-time language
the action
(\ref{Act}) takes the same form (\ref{commut}) but with the different
sign of the parameter $v$ and with the different sign of $\theta$ in
the Moyal product. In this formulation the gauge field is zero in the
$\bar{\mbox{D}}$-brane vacuum and non-zero in the D-brane vacuum. This
argument supports our claim that the noncommutativity parameter
changes its sign in the vacuum with $\phi=-v$.

Vacuum described by the direct sum of ${\cal N}$ D-brane
representations (\ref{Dbrane}) appears as a highly inhomogeneous state
in terms of the  $U(1)$ theory (\ref{commut}).
However, using the isomorphism between direct sum of ${\cal N}$ copies of the
Fock space and a single Fock space one may map this vacuum into the
vacuum with zero field in the $U({\cal N})$ NC gauge theory, whose
action is just a straightforward generalization of the action
(\ref{commut}) to the case of $U({\cal N})$ gauge group
(cf. Refs.~\cite{Gross:2000ss}-\cite{Demidov:2003xq},
\cite{Dubovsky:2003ga}).

Similar reinterpretation
of vacua where both D- and $\bar{\mbox{D}}$-branes are present is
somewhat more subtle. Indeed, we saw that for D-brane sector, it is
natural to use the space-time formulation with positive $\theta$,
while for $\bar{\mbox{D}}$-brane sector, the formulation
with negative value of $\theta$ is more appropriate. However, when
both sectors are present, there appear bifundamental fields
charged under gauge
groups of both sectors.
To understand what happens in this case,
it is instructive to study the physics of the D-$\bar{\mbox{D}}$ pair
(${\cal N}=1$, ${\cal M}=1$).

Let us use, for space-time interpretation, the formulation
with positive $\theta$. The action (\ref{Act})
then takes the form of the action of NC $U(2)$ gauge theory, which is
similar to (\ref{commut}) with the only difference that
all fields are now $2\times 2$ matrices and trace over $U(2)$ indices
is assumed. Background corresponding to D-$\bar{\mbox{D}}$
configuration has the following form (cf.
Eqs.~(\ref{vacrep})),
\be
\label{DaDbackgrnd}
\phi=
\begin{pmatrix}
v&0\\
0&-v
\end{pmatrix}~,~~~
X=
\begin{pmatrix}
q&0\\
0&-q
\end{pmatrix}~,~~~
Y=
\begin{pmatrix}
-i\theta\frac{\partial}{\partial q}&0\\
0&-i\theta\frac{\partial}{\partial q}
\end{pmatrix}\;.
\ee
It breaks $U(2)$ gauge group down to $U(1)\times U(1)$.
Magnetic field in this background is non-zero:
\be
\label{magnetic}
F_{xy}\equiv -\frac{1}{\theta^2}(i[X,Y]+\theta)=
\begin{pmatrix}
0&0\\
0&-\frac{2}{\theta}
\end{pmatrix}\;.
\ee
The block-diagonal structure of background (\ref{DaDbackgrnd})
ensures that spectra of diagonal
components of gauge fields and field $\phi$ are not modified and are
given by Eqs.~(\ref{dispAphi1}). On the other hand, off-diagonal
components feel both the VEV of the field $\phi$ and
the magnetic field (\ref{magnetic}). So, one expects
them to have Landau spectrum of energies with cyclotron frequency of
order $\frac{1}{\theta}$ and a constant piece of order $v^2$
(cf. Eq.~(\ref{gammadisp}) below). Instead of performing
explicit analysis of the gauge
sector, we confirm these statements
by working out the spectrum of a probe adjoint field $\chi$ in the
D-$\bar{\mbox{D}}$ background. This spectrum
shares essential features of the gauge field spectrum, while its
evaluation is more transparent.  

The field equation for the field $\chi$ reads
\be
\label{chieq}
-D_0^2\chi-\frac{1}{\theta^2}[X,[X,\chi]]
-\frac{1}{\theta^2}[Y,[Y,\chi]] -[\phi,[\phi,\chi]]=0\;.
\ee
Note that besides the interaction with gauge fields we introduce
interaction of the probe field with the scalar field $\phi$.
The latter interaction has the form typical for the adjoint fields
in the gauge theories residing on the stacks of D-branes, and
enables one
to interpret the field $\chi$ as the (matrix valued)
coordinate of the brane system in the extra transverse direction.
The diagonal structure of the background fields (\ref{DaDbackgrnd}) implies
that there is no mixing between different components
in $2\times 2$ matrix representation of
$\chi$,
\be
\label{chidecomp}
\chi=
\begin{pmatrix}
\alpha&\gamma\\
\gamma^+&\beta
\end{pmatrix}\;.
\ee
Equations for diagonal components $\alpha$ and $\beta$
following from (\ref{chieq}), (\ref{DaDbackgrnd})
coincide with the  equations for free massless scalar field
in the operator language. They have plane wave solutions
\[
\alpha,\beta\propto \e^{i(\omega t+k_x\hat{x}+k_y\hat{y})}
\]
with the massless dispersion relations
\be
\label{massless}
w^2=k_x^2+k_y^2.
\ee
In the string interpretation these components correspond to the strings with
both ends on one of the branes.

Let us now consider off-diagonal sector $\gamma$ which
corresponds to strings stretched between the brane
and the antibrane. It is convenient to consider $\gamma$ as an operator
in the space of functions of one variable $q$ (cf. Eqs.~(\ref{vacrep}))
and rewrite   Eq.~(\ref{chieq})
in terms of the kernel $\gamma(q,q')$,
\be
\label{gammaeq}
-\partial_0^2\gamma-\frac{(q+q')^2}{\theta^2}\gamma
+\left(\frac{\partial}{\partial q}+\frac{\partial}{\partial
    q'}\right)^2\gamma -4v^2\gamma=0\;.
\ee
Let us search for solutions of Eq.~(\ref{gammaeq}) of the
following form,
\be
\label{alphaeq}
\gamma(q,q')=\e^{i\omega t}\bar{\gamma}(\bar{q})\delta(u-\theta k_y)\;,
\ee
where $\bar{q}=(q'+q)/2$ and $u=q'-q$.
The $\delta$-function in Eq.~(\ref{alphaeq}) is the $q$-representation of
the operator $\e^{ik_y \hat{y}}$, so
Ansatz (\ref{alphaeq}) corresponds to the Fourier decomposition in
the $y$-direction. Then one arrives at the following equation for the
function
$\bar{\gamma}(\bar{q})$ \be
\label{bargammaeq}
\omega^2\bar{\gamma}=-\frac{\partial^2\bar{\gamma}}{\partial\bar{q}^2}
+\left(4v^2 +\frac{4\bar{q}^2}{\theta^2}\right)\bar{\gamma}\;.
\ee
Eq. (\ref{bargammaeq}) coincides with the eigenfunction equation for a
quantum mechanical oscillator. It yields the Landau spectrum,
\be
\label{gammadisp}
\omega^2=4v^2 +\frac{4}{\theta}\left(n+\frac{1}{2}\right)~,~~~n=0,1,2,\ldots
\ee
Note that for off-diagonal modes $\gamma$, the variable $u/2$ plays the
role of the $x$-coordinate of the string center, while $2\bar{q}$ is
the difference between coordinates of the ends of the string; this is
due to our choice of the background configuration (\ref{DaDbackgrnd}).
Consequently, the proper
interpretation of $k_y$ is that it determines the $x$-coordinate of
the excitation, namely
\be
\label{Landau}
x_0={k_y\theta\over 2}\;.
\ee
The constant piece $4v^2$
in the expression for energy
(\ref{gammadisp}) is generated by the VEV of the field $\phi$. In
the string language its presence is explained by the fact that the
minimal length of the string stretched
between brane and antibrane in the
D-$\bar{\mbox{D}}$ system is proportional to $v$.

Similar analysis applies to systems with larger number of branes.
The vacuum with  ${\cal N}$ branes and ${\cal M}$ antibranes
can be thought of as the trivial vacuum in the $U({\cal N})\times
U({\cal M})$ NC gauge theory with opposite signs of noncommutativity
parameter in different sectors and with extra bifundamental fields
interacting with both sectors. The latter fields exhibit the spectrum
of Landau type; they become
heavy and decouple in the commutative limit.

\section{Domain wall}
\label{WALL}
We turn now to the construction of a domain wall separating vacua with
different gauge groups. For concreteness let us  concentrate on
the
$U(1) - U(2)\times U(1)$ domain wall (on the one side of the wall the
gauge group is $U(1)$, while one the other side it is $U(2)\times U(1)$).
Energy functional for static field
configurations following from the action (\ref{Act}), can be
written as
\be
\label{potenergy}
\begin{split}
E=\frac{2\pi\theta}{g^2}\Tr\Bigl\{&\frac{1}{2\theta^2}(i[X,\phi])^2
+\frac{1}{2\theta^4}\left(i[X,Y]+\frac{\theta}{v}\phi\right)^2\\
&+\frac{1}{2}\left(\frac{i}{\theta}[Y,\phi]\pm\mu
  \left(\phi^2-v^2\right)\right)^2
\mp\frac{i\mu}{2\theta}\left[\phi^2Y+Y\phi^2-2v^2Y,\phi\right]\Bigr\}
\end{split}
\ee
This implies the BPS bound for energy,
\be
\label{BPSbound}
E\geq\frac{\mu v^2}{g^2}|Q|
\ee
where
\be
\label{topcharge}
Q=i\pi\Tr\left[\frac{\phi^2 Y+Y\phi^2}{v^2}-2Y,\phi\right]
\ee
is a topological charge (being a trace of a commutator, $Q$ does not
change under local variations of the fields).
Let us search for solutions saturating the bound
(\ref{BPSbound})
in the topological sector with a given charge $Q$. They obey
a system of BPS equations following from Eq.(\ref{potenergy}),
\bseq
\label{BPSeqs}
\begin{align}
\label{BPSeq1}
&[X,\phi]=0\\
\label{BPSeq2}
&i[X,Y]+\frac{\theta}{v}\phi=0\\
\label{BPSeq3}
&\frac{i}{\theta}[Y,\phi]\pm\mu (\phi^2-v^2)=0\;.
\end{align}
\eseq
These equations can be considered as defining an algebra with three
generators $X$, $Y$, $\phi$. Each solution
of Eqs.~(\ref{BPSeqs})
decomposes into a direct sum of operators acting in
irreducible representations of this algebra. Thus,
let us classify irreducible representations of the algebra
(\ref{BPSeqs}).

\begin{figure}[t]
\begin{center}
\epsfig{file=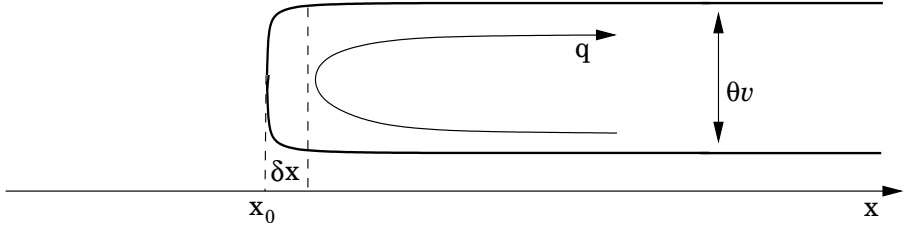,height=3.0cm,width=12.0cm}
\end{center}
\caption{Geometrical meaning of the parameters of the folded brane.}
\label{FOLD}
\end{figure}

Without loss of generality we work in the $x$-representation.
Namely, we assume that operators $X,Y,\phi$ act in the
Hilbert space of functions $\Psi(x)$ of a single variable $x$, and
\be
(X\Psi)(x)=x\Psi(x)\;.
\ee
From Eq. (\ref{BPSeq1}) it follows that
\be
(\phi\Psi)(x)=\phi(x)\Psi(x)\;,
\ee
where $\phi(x)$ is some yet unknown function. Eq. (\ref{BPSeq2})
yields
\be
(Y\Psi)(x)=-i\frac{\theta}{v}\phi(x)\frac{\partial\Psi (x)}{\partial x}\;.
\ee
Then Eq. (\ref{BPSeq3}) (where, to be specific, we
consider the upper sign) reads
\be
\label{phieq}
\frac{1}{v}\,\phi\,\frac{\partial\phi}{\partial x}+\mu (\phi^2-v^2)=0
\ee
This equation has three different bounded solutions:\\
{\em a}) $\phi=v$; in this case $-\infty<x<+\infty$,
$Y=-i\theta\frac{\partial}{\partial x}$, and, as it can be easily shown,
the topological charge $Q$ is zero. This solution corresponds to
a single $D$-brane located at the point $Z=\theta v$ in the transverse
direction.\\
{\em b}) $\phi=-v$; in this case $-\infty<x<+\infty$, and
$Y=i\theta\frac{\partial}{\partial x}$. The topological charge
$Q$ is zero as in the previous
case. This solution corresponds to an antibrane located at
$Z=-\theta v$.\\
{\em c}) $\phi=\pm v\sqrt{1-\e^{-2\mu v(x-x_0)}}$, $x_0\leq x$.
This solution describes a brane folded twice along the $x$-direction,
with the constant $x_0$ corresponding to the position of the tip of
the fold (see Fig.~\ref{FOLD}).
It is straightforward to check that $Q\neq 0$ in this case.
It is convenient to express this solution in
the $q$-representation, where the variable $q$ is a coordinate along the
fold; it is related to $x$ by
\be
dq=\frac{v\;dx}{\phi(x)}\;.
\ee
Straightforward calculation yields
\bseq
\label{foldq}
\begin{align}
\label{foldphi}
&\phi=\phi(q)\equiv v \tanh{(\mu v q)}\\
\label{foldX}
&X=x(q)\equiv\frac{1}{\mu v}\ln(2\ch(\mu v q)) +x_0\\
\label{foldY}
&Y=-i\theta\frac{\partial}{\partial q}\;.
\end{align}
\eseq
The geometrical meaning of the coordinates $x$ and $q$ is illustarted
in Fig.~\ref{FOLD}.
The width of the tip of the fold can be read off from
Eqs. (\ref{foldphi}), (\ref{foldX}),
\be
\label{wallwidth}
\delta x\approx \frac{1}{\mu v}\;.
\ee
Note that the condition (\ref{mutvcond}) implies that the distance
between two sheets of the folded brane
in the transverse direction, which is equal to
$\theta v$, is much larger than the width of the tip.

The $U(1) - U(2)\times U(1)$ domain wall is obtained by
taking the direct sum of representations ({\em a}) and ({\em c}),
\be
\label{DWbackgrnd}
\phi=
\begin{pmatrix}
v&0\\
0&\phi(q)
\end{pmatrix}~,~~~
X=
\begin{pmatrix}
q&0\\
0&x(q)
\end{pmatrix}~,~~~
Y=
\begin{pmatrix}
-i\theta\frac{\partial}{\partial q}&0\\
0&-i\theta\frac{\partial}{\partial q}
\end{pmatrix}\;,
\ee
where functions $\phi(q)$ and $x(q)$ are given by
Eqs. (\ref{foldphi}), (\ref{foldX}).
The solution (\ref{DWbackgrnd}) describes a wall
extending along $y$-direction,
which is located at the position $x_0$ on the $x$-axis.
To calculate the tension of the domain wall let us
plug the solution (\ref{DWbackgrnd}) into the expression
for the topological charge (\ref{topcharge}) and use
Eqs. (\ref{foldphi}), (\ref{foldY}). Then, from the BPS bound
(\ref{BPSbound}) one obtains for the energy
\be
\begin{split}
E&=-\frac{2\pi\theta\mu}{g^2}\Tr\left\{\frac{\partial\phi}{\partial q}
  \left(\phi^2-v^2\right)\right\}\\
&=-\frac{\mu}{g^2}\int dy\, dq\, \frac{\partial\phi}{\partial q}
  \left(\phi^2-v^2\right)
=\int dy\, \frac{4\mu v^3}{3g^2}\;,
\end{split}
\ee
where in the second equality we used the relation
$2\pi\theta\Tr = \int dy\,dq$. Thus, the tension of the wall
equals
\be
\label{walltension}
\sigma=\frac{4\mu v^3}{3g^2}\;.
\ee
In the commutative limit (\ref{commutlimit}) the width of the wall can
be kept 
finite, while its tension tends to infinity.

Construction presented above is straightforward to generalize to
the case of $U({\cal N})\times U({\cal M}) - U({\cal N}+{\cal
K})\times U({\cal M}+{\cal K})$
domain walls: one takes
direct sum of ${\cal N}$ representations of type ({\em a}), ${\cal M}$
representations of type ({\em b}) and ${\cal K}$ representations of
type ({\em c}). Let us also point out that though
in this section we have concentrated on domain walls extending along
$y$-axis, this direction is by no means preferred
from the point of view of the action (\ref{Act}). Performing rotation
of the solution constructed in this section, one can obtain
domain walls oriented arbitrarily in the $xy$-plane.

\section{Adjoint scalar in the domain wall background}
\label{Adjoint}
The purpose of this section is
to check the intuition coming from D-brane picture
and to confirm that the solution constructed in the previous section
describes a domain wall between regions with different gauge
theories. For concreteness,
we concentrate on the case of the $U(1)-U(2)\times U(1)$
domain wall. Generalization of our analysis to
$U({\cal N})\times U({\cal M}) - U({\cal N}+{\cal K})\times
U({\cal M}+{\cal K})$ domain wall
solutions is straightforward.

Again, as in section \ref{model},
for the sake of simplicity let us consider massless
adjoint scalar field $\chi$ as a probe and study its spectrum in the
domain wall background. Throughout this section we assume that
the condition (\ref{mutvcond}) is satisfied. This is the case in which our
results have the most transparent physical interpretation.

Field equation for the field $\chi$ is given by Eq.~(\ref{chieq}) with
background fields (\ref{DWbackgrnd}).
For definiteness we take $x_0=0$, {\it i.e.}
consider a domain wall with the tip at $x=0$. As in section
\ref{model}, we take $\chi$ in the form
(\ref{chidecomp}). Diagonal structure of (\ref{DWbackgrnd}) implies
that fields $\alpha$, $\beta$, $\gamma$ of this decomposition do not
mix.

Analysis of the spectrum in the $\alpha$-sector literally repeats that of
section \ref{model}. Thus, the field $\alpha$ describes massless particles
that are able to move freely from $x\to -\infty$ to $x\to +\infty$ and
do not feel the presence of the domain wall at all. These modes
describe excitations of
strings with both ends on the straight D-brane.

\begin{figure}[t]
\begin{center}
\begin{picture}(500,150)(0,0)
\put(0,20){\epsfig{file=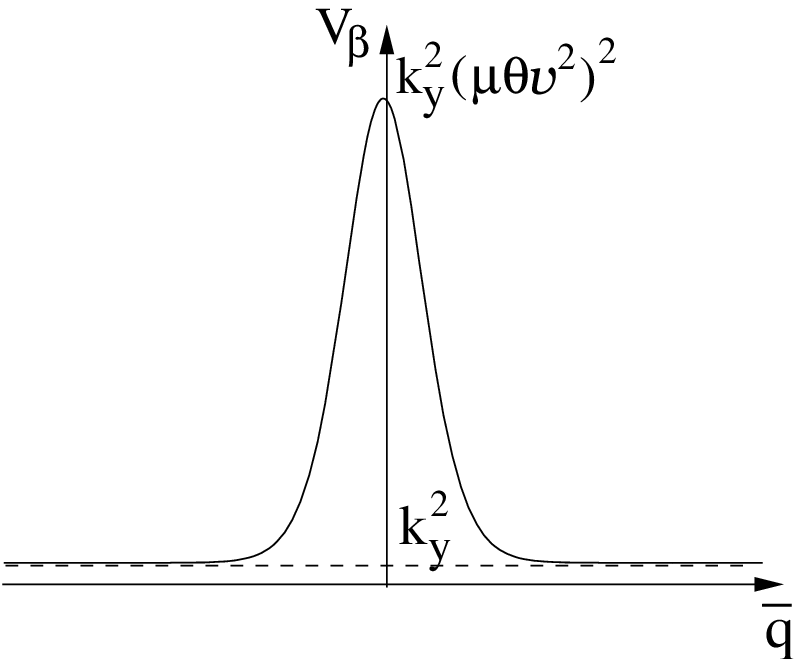,height=4.5cm,width=5.cm} }
\put(162,20){\epsfig{file=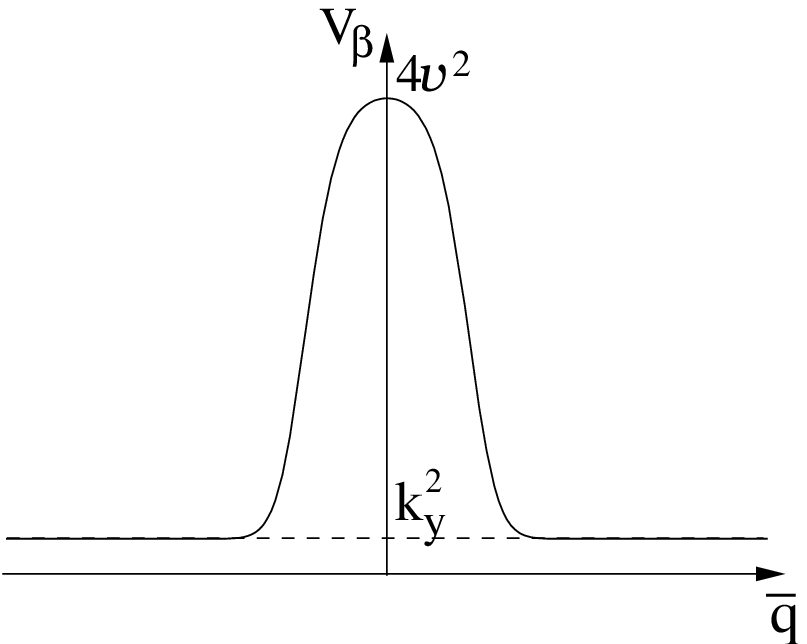,height=4.5cm,width=5.cm}}
\put(320,20){\epsfig{file=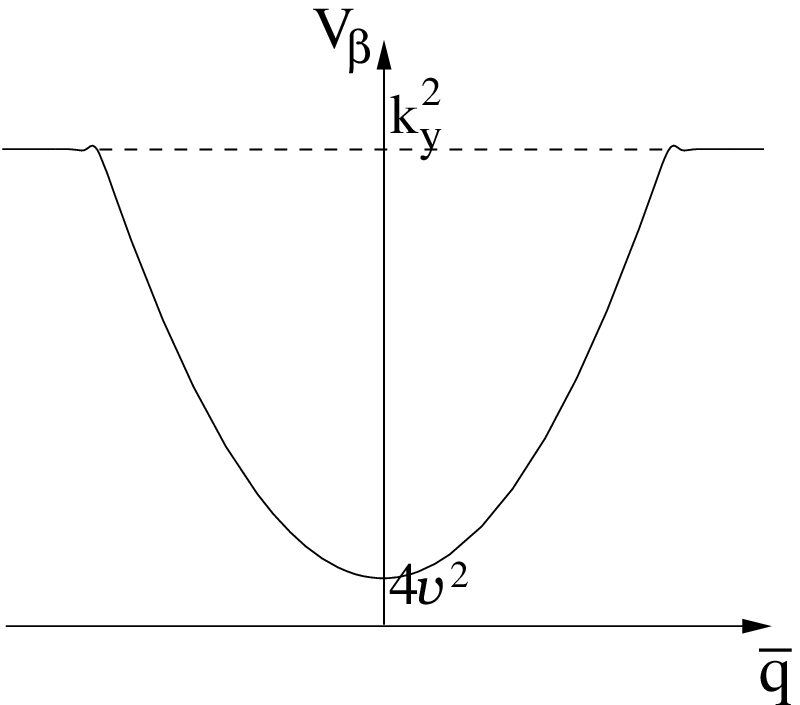,height=4.5cm,width=5.cm}}
\put(66,0){a}
\put(228,0){b}
\put(388,0){c}
\end{picture}
\end{center}
\caption{
Potential $V_{\beta}$ as a function of $\bar{q}$ for three regions of
parameter $k_y$:\newline
a) $k_y\ll\frac{1}{\mu\theta v}$~,~~ b) $\frac{1}{\mu\theta v}\ll k_y\ll
v$~,~~ c) $v\ll k_y$.}
\label{Vbetaplot}
\end{figure}

On the other hand, fields of the $\beta$-sector do not interact with
the straight brane, and live completely on the folded brane. From
Eqs.~(\ref{chieq}), (\ref{DWbackgrnd}) we get the following equation for
the kernel $\beta(q,q')$,
\be
-\partial_0^2\beta-\left(\frac{(x(q)-x(q'))^2}{\theta^2}
  +(\phi(q)-\phi(q'))^2\right)\beta +\left(\frac{\partial}{\partial
    q}+\frac{\partial}{\partial q'}\right)^2\beta=0\;.
\ee
Using the same Ansatz as in section \ref{model},
\be
\label{betasubs}
\beta(q,q')=\e^{i\omega t}\bar{\beta}(\bar{q})\delta(u-\theta k_y)\;,
\ee
where $\bar{q}=(q'+q)/2$, $u=q'-q$,
we arrive at the following
Schr\"{o}dinger-type equation for $\bar{\beta}$,
\be
\label{barbetaeq}
\omega^2\bar{\beta}=-\frac{\partial^2\bar{\beta}}{\partial\bar{q}^2}
+V_{\beta}(\bar{q},k_y)\bar{\beta}\;.
\ee
Here, the potential $V_{\beta}$ is given by
\be
\label{Vbeta}
V_{\beta}(\bar{q},k_y)=\left(\frac{1}{\mu v\theta}\ln{\frac{\ch\mu
      v\left(\bar{q}-\frac{\theta k_y}{2}\right)}{\ch\mu
      v\left(\bar{q}+\frac{\theta k_y}{2}\right)}}\right)^2
+v^2\left(\tanh\mu v\left(\bar{q}-\frac{\theta k_y}{2}\right)
  -\tanh\mu v\left(\bar{q}+\frac{\theta k_y}{2}\right)\right)^2\;.
\ee
This potential is shown in Fig. \ref{Vbetaplot} for three
different regions of parameter $k_y$.
In all three cases it approaches $k_y^2$ as
$\bar{q}\to\pm\infty$. Thus, asymptotically, the solution to
Eq.~(\ref{barbetaeq}) is given
by plane waves $\bar{\beta}\propto \e^{\pm ik_x \bar{q}}$ with
dispersion relation (\ref{massless}). Two regions, $(\bar{q}>0)$
and $(\bar{q}<0)$ correspond to the upper and lower sheets of the
folded brane,
respectively. We will refer to
modes of the field $\beta$ living on the upper (lower) sheet as $\beta_+$
($\beta_-$) sector. This terminology
does not imply that $\beta_+$ and
$\beta_-$ sectors are disconnected: as we will see immediately there
are modes that can
propagate round the tip of the folded brane from one sheet to another.

The behavior of the
potential $V_{\beta}$ in the region near the tip, $\bar{q}\approx
0$, depends on the value of
$k_y$. For
$k_y\ll\frac{1}{\mu\theta v}$ (see Fig. \ref{Vbetaplot}a) it has a sharp
peak around
$\bar{q}=0$ with the height $k_y^2(\mu\theta v^2)^2\gg k_y^2$. This
means that modes with small momentum along $x$-axis
are reflected backwards from the tip,
their propagation through the tip being suppressed. However, there are
modes that can propagate round the tip without suppression, namely,
those whose wave-vector is almost perpendicular to the tip,
$k_x^2>k_y^2(\mu\theta v^2)^2$.

When $\frac{1}{\mu\theta v}\ll k_y\ll v$, the potential barrier
separating $\beta_+$ and $\beta_-$ sectors is
still present, but now its height is equal to $4v^2$
(see Fig. \ref{Vbetaplot}b).

Behavior of $V_{\beta}(\bar q)$ is qualitatively different for $v\ll
k_y$ (Fig. \ref{Vbetaplot}c). The potential has a dip at the origin
with minimal
value $4v^2$. Around
this minimum it can be approximated as follows,
\be
V_{\beta}\approx 4v^2+\frac{4\bar q^2}{\theta^2}\;.
\ee
Thus the energy spectrum for low-lying modes is
given by Eq.~(\ref{gammadisp}) in this case.
We will refer to these modes as $\beta_*$-sector.

\begin{figure}[t]
\begin{center}
\begin{picture}(500,100)(0,0)
\put(0,25){\epsfig{file=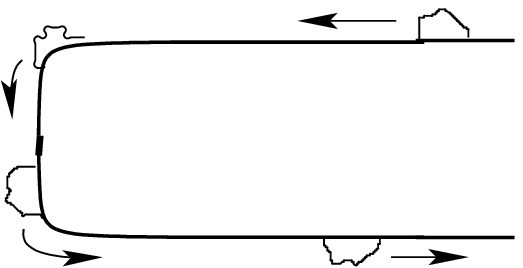,height=2.05cm,width=4.5cm} }
\put(162,20){\epsfig{file=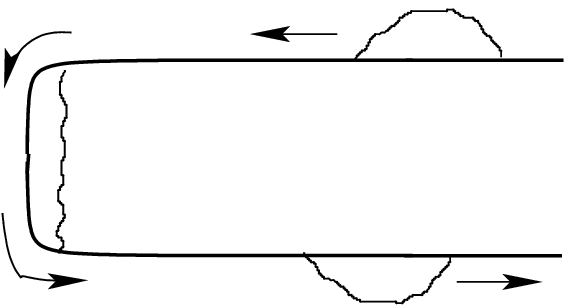,height=2.5cm,width=4.5cm}}
\put(320,33){\epsfig{file=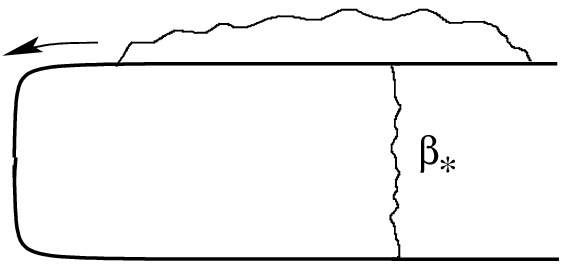,height=2.1cm,width=4.5cm}}
\put(66,0){a}
\put(228,0){b}
\put(388,0){c}
\end{picture}
\end{center}
\caption{
Stringy interpretation of small perturbations in the
folded brane background. \newline
a) $k_y\ll\frac{1}{\mu\theta v}$~,~~ b) $\frac{1}{\mu\theta v}\ll k_y\ll
v$~,~~ c) $v\ll k_y$.
}
\label{Strings}
\end{figure}

Many of these features have physical explanation in the dipole
interpretation~\cite{Sheikh-Jabbari:1999vm,Bigatti:1999iz}.
Namely, one may consider excitations in
the NC theory as dipoles (strings) whose length in, say, $q$-direction
is proportional to the momentum in the perpendicular $y$-direction,
\be
\label{dipole}
\Delta q=\theta k_y\;.
\ee
Then in the first kinematical region, $k_y\ll1/(\mu\theta v)$,
the length of the string in $q$ direction is
much smaller than the characteristic curvature radius of the
tip.
Both ends of such string enters into the curved region simultaneously,
see Fig. \ref{Strings}a.
The fact that the height of the potential barrier grows with $k_y$ is easy
to understand --- the longer the string is, the more difficult it is
for this string
to turn round the tip. It would be interesting to calculate the
coefficient of proportionality between $k_y$ and the height of the
barrier in the string language.

In the second region, the situation changes. Here the length of the string
is larger than the curvature radius of the tip.
The ends of such string do not
turn round the tip at one time anymore. Instead, first
the left end turns round the tip, and only after it reaches the second
sheet, the right end follows, see Fig.~\ref{Strings}b.
In this regime the height of the
barrier does not depend on $k_y$. Rather, it is given by the energy
of the string stretched between the two sheets, which is equal to $2v$.

Finally, the third region corresponds to strings with separation between
their ends in the
$q$-direction much larger then the distance between the two sheets of the
fold.
Here, two types of strings exist. First, there are very long strings
that do not experience any potential barrier while propagating from
one sheet
to another. At the same time new type of states
appear ($\beta_*$-sector). These are strings stretched between different
sheets of the folded brane, see Fig.~\ref{Strings}c.
They correspond to the heavy states
(\ref{gammadisp}) present in the spectrum of D-$\bar{\mbox{D}}$
system. Note that for these strings, $k_y$ measures the distance from
the tip of the fold. Here
the difference between $q$ and $x$
coordinates is crucial. Namely, for these strings the difference between $q$
coordinates of their ends is still given by
Eq.~(\ref{dipole}), so these strings exist only far enough
from the tip. However, as $q$ and $-q$ correspond to the same
value of the physical $x$ coordinate, the actual length of these
 strings can be as small as the distance $2\theta v$ between the
two sheets of the folded brane.

To summarize, far from the tip,
 we see that the spectrum of the folded brane reproduces the spectrum
of the D-$\bar{\mbox{D}}$
system.

\begin{figure}[t]
\begin{center}
\begin{picture}(500,150)(0,0)
\put(0,20){\epsfig{file=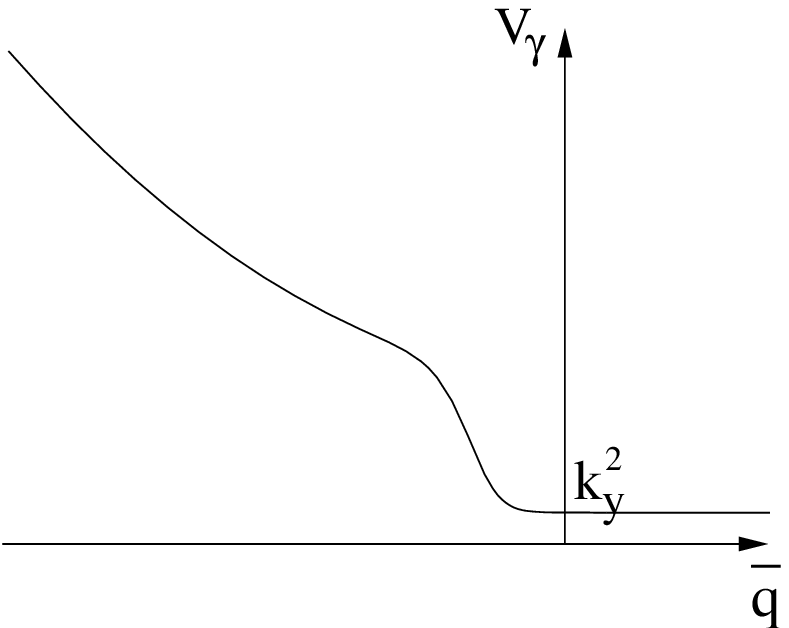,height=4.5cm,width=7.cm} }
\put(250,20){\epsfig{file=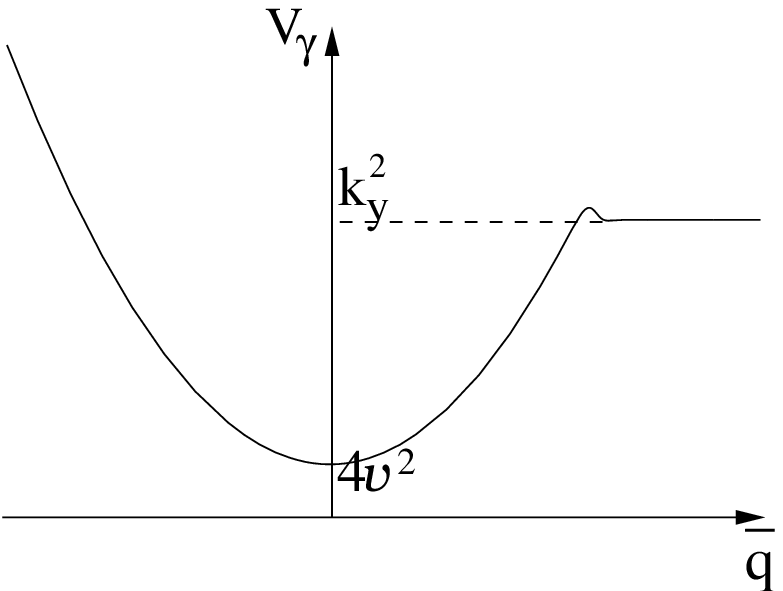,height=4.5cm,width=7.cm}}
\put(95,0){a}
\put(345,0){b}
\end{picture}
\end{center}
\caption{
Potential $V_{\gamma}$ as a function of $\bar{q}$ for two regions of
parameter $k_y$:\newline
a) $k_y>-v$~,~~ b) $k_y<0,~~|k_y|\gg v$.}
\label{Vgammaplot}
\end{figure}

Let us now consider the $\gamma$-component of the field $\chi$ (see
Eq. (\ref{chidecomp})). This component corresponds to strings with one
end on the
straight brane and the other on the folded one. Here we have the
following equation
for the kernel $\gamma(q,q')$,
\be
-\partial_0^2\gamma
-\left(\frac{(q-x(q'))^2}{\theta^2}+(v-\phi(q'))^2\right)\gamma
+\left(\frac{\partial}{\partial q}+\frac{\partial}{\partial
    q'}\right)^2 \gamma=0\;.
\ee
Using Ansatz analogous to (\ref{betasubs}) we get
\be
\omega^2\bar\gamma=-\frac{\partial^2\bar\gamma}{\partial \bar q^2}
+V_{\gamma}(\bar q, k_y)\bar\gamma\;,
\ee
where
\be
\label{Vgamma}
\begin{split}
V_{\gamma}(\bar q, k_y)=\frac{1}{\theta^2} \left\{\bar q-\frac{\theta
    k_y}{2}-\frac{1}{\mu v}\ln\left( 2\ch\mu v\left(\bar q+\frac{\theta
    k_y}{2}\right)\right)\right\}^2~~~~~~~~~~~~~~~~~~~& \\
+v^2\left(1-\tanh\mu v\left(\bar
    q+\frac{\theta k_y}{2}\right)\right)^2&\;.
\end{split}
\ee
Qualitative behavior of the potential
$V_{\gamma}$ in two different regions of the parameter
$k_y$ is shown in
Fig. \ref{Vgammaplot}.
In both cases its asymptotics are
\begin{gather}
V_{\gamma}\to k_y^2~,~~~\bar q\to +\infty\\
V_{\gamma}\sim\frac{4\bar q^2}{\theta^2}~,~~~\bar q\to -\infty
\end{gather}
Thus, $\gamma$-sector contains a sector of plane waves living in the
region $\bar q>0$. Let us call these modes $\gamma_+$-sector. They correspond
to strings with one end on the straight brane and the other on the
upper sheet of the
folded brane.
One of the ends of these strings always has positive
$x$-coordinate. This explains why the  $\gamma_+$-modes cannot penetrate
into the region of negative $\bar q$, which corresponds to negative
$x$-coordinate on the straight brane.

When $k_y<0,~~|k_y|\gg v$, the potential $V_{\gamma}$ has a dip near the
origin (Fig.~\ref{Vgammaplot}b) similar to the case of $\beta$-sector
(cf. Fig.~\ref{Vbetaplot}c). In the vicinity of the minimum one has
\be
V_{\gamma}\approx 4v^2+\frac{4\bar q^2}{\theta^2}\;.
\ee
The energy spectrum of low-lying modes in this case is again given by
Eq. (\ref{gammadisp}). These modes correspond to strings stretched
between the straight brane and the lower sheet of the folded one. We
denote this sector by $\gamma_*$.

Let us summarize the results of this section and give an interpretation
to the spectrum of the adjoint field $\chi$ from the viewpoint of
 $(2+1)$ dimensional field theory. In the region $x\to -\infty$
there is one massless field $\alpha$. This means that in this region
we have $U(1)$ gauge theory. In the region $x\to +\infty$ this field,
 together with $\beta_+$-, $\gamma_+$-fields, form the  adjoint multiplet
of the $U(2)$ gauge group
\be
\begin{pmatrix}
 \alpha&\gamma_+\\
 \gamma_+^+&\beta_+
\end{pmatrix}
\ee
Note that the field $\alpha$ propagates freely from $x=-\infty$ to
$x=+\infty$. Thus, the $U(1)$ and $U(2)$ gauge theories on either side
of the wall are connected even at low energies. In addition
there is another
massless field $\beta_-$ at $x\to +\infty$. This is a
signal that an additional $U(1)$ gauge group is present in this
region.
Let us stress that this $U(1)$
gauge group is different from the $U(1)$ in the region $x\to
-\infty$.
Finally, there are high-energy modes $\beta_*$, $\gamma_*$ with the
Landau spectrum (\ref{gammadisp}) living at large enough positive $x\gtrsim\theta v$.

These results
confirm our statement, that the solution obtained in the previous section
describes a domain wall between regions with $U(1)$ and $U(2)\times
U(1)$ gauge theories.

\section{Conclusions and discussion}
\label{final}
From the field
theory viewpoint we presented an example of a gauge theory with adjoint
scalar on NC plane, such that the rank and the type of the gauge group
in it are not fixed, but can change dynamically. In particular, they
can vary along the plane. We constructed explicit
solutions in the form of domain walls interpolating between regions in
space with different gauge groups. Though our model is formulated in
$(2+1)$ dimensions, it can be easily generalized to higher dimensional
space-time by adding an arbitrary number of commuting space-like
dimensions. Then, the higher dimensional model possesses domain wall
solutions independent of the commuting coordinates.

The fact that the gauge groups on different sides of the wall are
different was
confirmed by explicit evaluation of the spectrum of a probe adjoint
field. Namely, we demostrated that its excitations with equal momentum and
energy form adjoint multiplets of different gauge theories on the two
sides of the wall. As in the case of domain walls on the fuzzy
cylinder \cite{Dubovsky:2003ga} a similar analysis can be performed
for fundamental probe fields and for the gauge fields themselves. An
outcome of this analysis is that there are always modes charged under
the gauge group, that can freely propagate through the wall. This
holds even for the low-energy part of the spectrum whose dynamics on
either side
of the wall is
desribed (barring potential UV/IR mixing effects)
by commutative gauge theory. These results
indicate that (non)commutative $U({\cal N})$ gauge
theories may be thought of as {\em dynamically connected} sectors of a
single theory. It is interesting to explore
this relationship at the quantum level. We leave this investigation for
future studies.

We have seen above that many properties of the
model considered in this paper have natural physical interpretation in
stringy
terms. This suggests that there should be a way to embed domain walls
discussed in this paper
into the context of string/M-theory. Let us
briefly present arguments indicating
that such embedding is quite plausible.
These arguments are
parallel to those presented in the
case of the domain wall on the fuzzy cylinder~\cite{Dubovsky:2003ga}.
We refer an interested
reader to Ref.~\cite{Dubovsky:2003ga} for more details.

Let us recall that according to the BFSS conjecture \cite{BFSS}, the
dynamics of the M-theory in the infinite momentum frame is described
by the large-$N$ limit of the BFSS Matrix model. The latter is
supersymmetric quantum mechanics described by the following
Lagrangian (see,
e.g., Ref. \cite{Taylor} for a review)
\be
\label{1*}
L={1\over 2R}\Tr\left\{ \dot{ X}^i\dot{ X}^i+{1\over
2}[{ X}^i,{ X}^j]^2+(\mbox{fermions})\right\}\;,
\ee
where ${ X}^i$ ($i=1,\dots,9$) are real-valued $N\times N$ matrices
subject to the constraint
\[
[\dot{ X}^i,{ X}^i]=0\;.
\]
The matrices ${ X}^i$ are (matrix-valued)
coordinates of D0 branes (in string units $l_s=1$) which are
conjectured to be the partons of M-theory.
The parameter $R$ is the string coupling constant which is interpreted
as the  compactification radius of 11-dimensional M-theory
to ten dimensions.

Here we would like to suggest that domain walls studied in this paper
may be obtained  as solutions of the Matrix model in {\cal curved}
backgrounds.
Our proposal is based on the observation that
if one sets
${ X}^i=0$ for $i\geq 4$
in the Matrix model Lagrangian, one arrives at the action very
similar to the action (\ref{Act}) where $X,Y$ and $Z=\theta\phi$
correspond to matrix coordinates $X_1,\;X_2,\;X_3$, string
length $l_s$ is equal to the scale of noncommutativity,
$l_s=\theta^{1/2}$ and $R=g^2\theta$. The only difference is that
in our theory we have a
potential for the field $\phi$ and that extra $\phi$-dependence is
present in the first
term $\frac{\theta}{v}\phi$ in the second line of Eq.~(\ref{Act}).
Terms of this structure are expected to appear
in the Lagrangians of Matrix models corresponding to a large class of
curved backgrounds of M-theory \cite{Taylor}.

It is worth noting, that generalization of the Matrix model
Lagrangian (\ref{1*}) to arbitrary curved background is not known
(see Ref.~\cite{Douglas:1997ch} for a discussion of this problem). However,
there is a proposal \cite{Taylor:1999gq} on how to modify the
Lagrangian of the Matrix
model to incorporate the effect of arbitrary  weakly curved
background independent of the light
cone coordinate $x^-$.
This proposal makes it natural to consider~\cite{Dubovsky:2003ga} the
metric of the eleven-dimensional pp-wave
\begin{align}
\label{pp}
&ds^2=-2dx^+dx^-+\sum dx^ldx^l-H(x^l)(dx^+)^2\;.
\end{align}
In order to be a legitimate background of the M-theory, at least in
the supergravity approximation, this metric should be
supplemented by
an appropriate three-form
field to satisfy the equations of eleven-dimensional
supergravity. This class of solutions was found in Ref.~\cite{Hull:vh}
and homogeneous pp-wave considered recently in Ref.~\cite{Berenstein:2002jq}
is  a particular example of the background from this class.

Then, following the prescription
of Ref.~\cite{Taylor:1999gq},
it is straightforward to find various backgrounds in which
the action of the BFSS Matrix model contains the terms
present in Eq.~(\ref{Act}).
For instance, the metric of the form (\ref{pp}) with the
function
\be
H=\frac{1}{a^2}z^2+\mu^2(z^2-a^2)^2+2\mu^2a^2x_4^2+\mu^2x_4^4
\ee
supplemented with the following three-form
\be
A=\frac{1}{a}(z dx_+\wedge dx\wedge dy+x_4dx_+\wedge dx\wedge
dy)+
\sqrt{12}\mu zx_4dx_+\wedge dx_5\wedge dx_6
\ee
will do the job\footnote{Note, that unlike in the previous section, we
are working in the string units $\theta=l_s^2=1$.}.
It is worth noting that with this choice of
background some extra terms besides those of Eq.~(\ref{Act})
will appear in the Lagrangian of the
Matrix model. However, these terms are at least quadratic in the
coordinates $X_i$ with $i>3$ and positive definite at the quadratic
level.
Consequently, they affect neither existence nor perturbative
stability of the brane configurations considered above.
Thus, it is natural to expect that configurations of this
type naturally appear in a large class of curved backgrounds.

Clearly, our arguments rely on the approximation of weakly
curved background; one may hope that they apply beyond this
approximation, especially taking into account that pp-wave
backgrounds similar to those we discussed here were
shown to be exact string backgrounds
\cite{Russo:2002qj,Bonelli:2002fs}.
We leave aside an issue of the supersymmetrization of the domain walls,
though the BPS property suggests that it should be possible.

Finally, note that the folded brane
involved into construction of the domain wall can be considered by
itself and provides an explicit
realization of the old idea of ``mirror world'' \cite{LY} (for recent
review see Ref.~\cite{Berezhiani:2003xm}). Similar idea was put forward
in Ref.~\cite{Arkani-Hamed:1999zg} in the context of brane world
models under the name of ``many fold Universe''.
According to this scenario our world is a brane folded several
times inside the bulk space.  Matter residing on the different sheets of
the folded brane is microphysically identical to the ordinary matter
(in our case these two types of matter are represented by $\beta_+$
and $\beta_-$ sectors), but
at low energies only gravitational interaction is possible between
them.
As a result, matter on a sheet different from ours
may be observed as dark matter.

Some possible phenomenological
aspects of this scenario were considered in
Ref.~\cite{Arkani-Hamed:1999zg}. Here we would like to mention just
one more. As we saw in
section \ref{Adjoint}, there are high energy modes ($\gamma_*$ and $\beta_*$)
corresponding to strings stretched between different sheets of the
folded brane. At high enough energy, production of these modes and,
as a result, direct connection between different sheets is possible.\footnote{Note 
that in our model the dispersion relation for these modes is
highly Lorentz-noninvariant. It is interesting to understand whether
this is a generic phenomenon and what are the phenomenological
implications of this violation of Lorentz invariance.}

With a bit of fantasy one
may envisage an exciting possibility that an intelligent life
exists in the Solar system, but on a different sheet of our brane. Then,
creating strings stretched between the two sheets of the brane at
future accelerators, such as LHC\footnote{LHC= Large Humanoid
Communicator.},
one can establish direct communication with this civilization,
with information transfer in the form, say, of the Morse code.
Equally
exciting is to look for messages from this civilization in the
ultra-high energy cosmic ray data. Clearly, this possibility appears
extremely unlikely. The only argument we may present in favor of this
idea is due to Freeman Dyson~\cite{dyson}: ''astronomy teaches us that
we should look not for those things which are likely, but for those
which are detectable''.

\section*{Acknowlegements}
We are indebted to D.~Gorbunov and V.~Rubakov for useful discussions.
This work has been supported in part by Russian Foundation for Basic
Research, grant 02-02-17398, and the grant of the President
of Russian Federation NS-2184.2003.2. The work of S.S. has been supported in
part by INTAS grant YS 2001-2/141. S.S. is grateful
to DESY Theory Group in Hamburg and Institute of Theoretical Physics,
EPFL, Lausanne for hospitality during his visits
to these organizations.

\end{document}